\documentclass[amssymb,amsmath,preprint,nofootinbib]{revtex4-2}
\usepackage[colorlinks]{hyperref}
\usepackage{graphicx}
\usepackage{subcaption}
\usepackage{float}
\usepackage{xcolor}

\begin{document}

\title{Search for Axion Like Particles\\produced via the Primakoff process at COMPASS}

\author{Mehran Dehpour}
\email{m.dehpour@cern.ch}
\affiliation{Faculty of Mathematics and Physics, Charles University, Prague, Czechia}

\begin{abstract}
    Axion-Like Particles (ALPs) are well-motivated candidates for dark matter and potential mediators to the dark sector.
    We present a search for ALPs coupled to photons, based on a reinterpretation of COMPASS data.
    Using the 2009 dataset consisting of $190~\text{GeV}$ $\pi^-$ and $\mu^-$ beams impinging on a fixed nickel target, we investigate the Primakoff production of ALPs.
    Due to the high beam energy, ALPs in the MeV mass range are produced with a significant Lorentz boost, leading to strongly collimated decay photons.
    Consequently, these photons are not spatially resolved by the electromagnetic calorimeter and are instead reconstructed as a single merged cluster.
    This signature mimics the single-photon signal of Standard Model Primakoff Compton scattering, which was the primary focus of the original COMPASS analysis.
    By quantifying this potential ALP contamination in the Compton scattering sample, we derive exclusion limits on the ALP-photon coupling $g_{a\gamma\gamma}$ in the mass range $0.2 \lesssim m_a \lesssim 600~\text{MeV}$.
    Our results exclude couplings $g_{a\gamma\gamma} \gtrsim 10^{-1}~\text{GeV}^{-1}$ at 95\%~C.L., providing independent constraints on the parameter space that bridges beam dump experiments and high energy colliders.
    While current collider-based limits remain more stringent, this work establishes a novel reinterpretation framework and provides a baseline for future studies of resolved diphoton states in complementary kinematic regimes, such as Primakoff $\pi^0$ production.
\end{abstract}

\maketitle

\clearpage

\section{Introduction}
\label{sect:intro}
The Standard Model (SM) of particle physics has proven remarkably successful in describing fundamental interactions.
However, the strong interaction sector presents a long-standing theoretical puzzle known as the strong $CP$ problem.
The Quantum Chromodynamics (QCD) Lagrangian naturally admits a $CP$-violating term, proportional to the $\theta$-vacuum angle, which should theoretically induce a large electric dipole moment for the neutron.
Yet, experimental bounds on the neutron electric dipole moment constrain this term to a value unnaturally close to zero~\cite{Abel:2020pzs}, implying a fine-tuning problem of nearly ten orders of magnitude.
This issue is elegantly addressed by the Peccei-Quinn mechanism~\cite{Peccei:1977hh}, which postulates a global chiral $U(1)$ symmetry.
The spontaneous breaking of this symmetry at a high energy scale yields a pseudo-Nambu-Goldstone boson known as the axion~\cite{Weinberg:1977ma,Wilczek:1977pj}.

While the properties of the canonical QCD axion are strictly determined by the relationship between its mass and decay constant, extensions of the SM predict a broader class of Axion-Like Particles (ALPs).
ALPs share the generic property of coupling to gauge bosons, particularly photons, but inhabit a wider parameter space unconstrained by the strict mass-coupling relations of the original QCD model.
Such particles appear naturally in string theory~\cite{Svrcek:2006yi,Arvanitaki:2009fg} and supersymmetric extensions of the SM~\cite{Nelson:1993nf}.
Furthermore, ALPs are viable candidates for dark matter or mediators to the dark sector~\cite{Preskill:1982cy,Abbott:1982af,Dine:1982ah}, motivating a diverse experimental program to search for them~\cite{ParticleDataGroup:2024cfk}.
In recent years, significant efforts have been made to constrain ALPs in the MeV to GeV mass range.
Notably, reinterpretations of LEP data~\cite{Jaeckel:2015jla,Knapen:2016moh} have placed stringent bounds on ALP production at higher masses.
Conversely, beam dump experiments~\cite{ParticleDataGroup:2024cfk} are typically limited to lower masses and weaker couplings due to the requirement of a macroscopic decay length.

The COMPASS experiment at the CERN SPS~\cite{COMPASS:2007rjf} provides a unique environment to probe the intermediate transition region via the Primakoff mechanism using high energy particle beams on a nuclear target.
In 2009, COMPASS recorded a dedicated dataset using $\pi^{-}$ and $\mu^{-}$ beams with a momentum of $190~\text{GeV}$ on a fixed nickel target~\cite{COMPASS:2014eqi}.
The primary motivation for this campaign was the precise measurement of the charged pion polarizability via pion-induced Primakoff Compton scattering, $\pi^{-}\text{Ni} \to \pi^{-}\text{Ni}\gamma$.
A parallel analysis using a muon beam was performed to validate the experimental methodology, exploiting the point-like nature of the muon to verify the expected QED cross section and detector response.

This work proposes a novel reinterpretation of the COMPASS data by investigating the Primakoff production of ALPs, $\pi^{-}\text{Ni} \to \pi^{-}\text{Ni}a$ and $\mu^{-}\text{Ni} \to \mu^{-}\text{Ni}a$.
Assuming the ALP couples predominantly to photons, it decays into a diphoton state.
Due to the high energy of the incident beams, the produced ALPs are highly boosted, leading to a strongly collimated photon pair.
Given the flight distance of approximately $34~\text{m}$ between the target and the electromagnetic calorimeter (ECAL2), combined with its $3.8~\text{cm}$ cell granularity, the transverse separation of these photons can be smaller than the resolving power of the detector.
In this regime, the resulting electromagnetic showers merge into a single reconstructed cluster, perfectly mimicking the single-photon signature of SM Primakoff Compton scattering.
By quantifying this potential contamination using an analytical model for cluster merging based on the ECAL2 cellular response, we derive exclusion limits on the ALP-photon coupling.

This analysis provides a direct and independent test of the ALP-photon sector in the mass range $0.2 \lesssim m_a \lesssim 600~\text{MeV}$.
Our results exclude couplings $g_{a\gamma\gamma} \gtrsim 10^{-1}~\text{GeV}^{-1}$ at 95\%~C.L., providing complementary constraints spanning the intermediate mass region between beam dump experiments and high energy colliders~\cite{ParticleDataGroup:2024cfk}.
This work complements recent results from other Primakoff-based experiments, such as PrimEx~\cite{Aloni:2019ruo} and GlueX~\cite{Pybus:2023yex}.
However, our reinterpretation offers a distinct experimental advantage. Photoproduction experiments rely on real photon beams, making the identification of highly collimated (merged) diphoton signatures exceptionally challenging due to the overwhelming background from the non-interacting beam. In contrast, the COMPASS setup utilizes charged particle beams. The detection of the scattered charged particle provides a clean coincidence tag for the interaction vertex. This crucial topological feature naturally suppresses beam-related backgrounds and makes the search for unresolved single-cluster topologies feasible.
Furthermore, it establishes a methodological framework for the future analysis of the high-statistics COMPASS 2012 dataset and serves as a baseline for searches focusing on resolved diphoton states in Primakoff $\pi^0$ production.

The paper is organized as follows. Section~\ref{sect:compton} reviews the theoretical framework of Primakoff Compton scattering, followed by a detailed description of the ALP production mechanism in Sec.~\ref{sect:alp}.
The analysis strategy and reinterpretation methodology are discussed in Sec.~\ref{sect:analysis}, and finally, Sec.~\ref{sect:results} presents the exclusion results and conclusions.

\section{Primakoff Compton scattering}
\label{sect:compton}
\begin{figure}[h]
    \centering
    \begin{minipage}{0.45\textwidth}
        \centering
        \includegraphics[page=1]{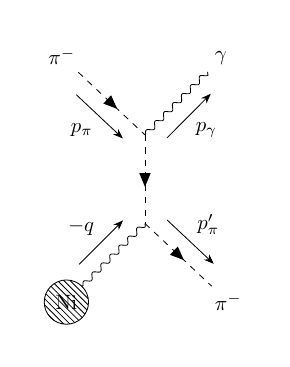}
    \end{minipage}
    \begin{minipage}{0.45\textwidth}
        \centering
        \includegraphics[page=2]{compton.pdf}
    \end{minipage}
    \caption{Primakoff Compton scattering of a $\pi^-$ on a Ni, with implicit coherent recoil}
    \label{fig:compton}
\end{figure}

As discussed in Sec.~\ref{sect:intro}, the original COMPASS analysis investigated SM Primakoff Compton scattering of $\pi^-$ and $\mu^-$ beams off nickel nuclei.
The choice of a high-$Z$ target exploits the coherent $Z^2$ enhancement of the electromagnetic interaction to effectively suppress the relative contribution of hadronic background processes.
Figure~\ref{fig:compton} illustrates the schematic tree-level Feynman diagrams for the pion-induced process.
In these diagrams, since the nucleus recoils coherently, its outgoing line is omitted for clarity.
The corresponding muon-induced process proceeds identically, by simply substituting the pion lines with muons.

\subsection{Muon bremsstrahlung on nucleus}
For the muon case, the scattering process corresponds to bremsstrahlung off a nucleus.
This can be described analytically in the Born approximation, yielding the well-known Bethe-Heitler (BH) formula~\cite{Bethe:1934za}.
Alternatively, one can describe it via the Weizsacker-Williams method~\cite{vonWeizsacker:1934nji,Williams:1934ad}, which approximates the electromagnetic field of the nucleus as a spectrum of equivalent photons, known as the Equivalent Photon Approximation (EPA).
This allows the bremsstrahlung cross section to be expressed as the muon Compton cross section convolved with the nuclear equivalent photon flux.
The two approaches yield consistent results in the relevant kinematic regime (see, e.g., Ref.~\cite{Kim:1973he}).
The resulting BH expression for the differential cross section of muon bremsstrahlung off a target nucleus with atomic number $Z$ is given by:
\begin{align}
    \frac{{\rm d}\sigma_\mu^\text{Born}}{{\rm d}E_\gamma}=\frac{4 Z^2 \alpha^3}{m_\mu^2}\frac{E_\mu'}{E_\mu E_\gamma}\left(\frac{E_\mu}{E_\mu'}+\frac{E_\mu'}{E_\mu}-\frac{2}{3}\right)\left[\ln\frac{|\vec{q}_\text{max}|}{|\vec{q}_\text{min}|}-\frac{1}{2}\right],
\end{align}
where $\alpha$ is the fine-structure constant, $E_\mu$ and $E_\mu'$ are the energies of the incoming and scattered muon, respectively, $E_\gamma$ is the energy of the outgoing real photon, and $\vec{q}$ is the three-momentum transfer to the nucleus.
The minimum momentum transfer is kinematically defined as:
\begin{align*}
    |\vec{q}_\text{min}|=\frac{E_\gamma m_\mu^2}{2 E_\mu E_\mu'},
\end{align*}
while the upper bound, $|\vec{q}_\text{max}|^2=0.0015~\text{GeV}^2$, is determined by the experimental analysis selection to suppress contributions from the strong interaction~\cite{COMPASS:2014eqi}.

This expression is derived assuming a point-like nucleus, thereby neglecting atomic screening and nuclear form factors~\cite{Andreev:1997pf}.
This point-like approximation is well-justified for modeling all processes in our analysis.
The experimental selection strictly limits the momentum transfer, ensuring that the scattering remains entirely within the coherent regime where the internal structure of the nickel nucleus is completely unresolved.
Furthermore, since the SM background and ALP signal would be modified similarly, any residual nuclear form factor effects cancel to a very good approximation in the cross section ratio introduced in Sec.~\ref{sect:analysis}.

\subsection{Charged pion bremsstrahlung on nucleus}
\label{sect:pion}
Analogous to the muon case, the bremsstrahlung of a charged pion off a nucleus has been calculated using both the Born approximation and the EPA in Ref.~\cite{Galperin:1980af}.
The validity of this calculation relies on the kinematic condition $|\vec{q}_\text{min}|^2 \ll |\vec{q}_\text{max}|^2 \ll m_\pi^2$. This requirement is satisfied in our analysis, given the experimental selection cut of $|\vec{q}_\text{max}|^2 = 0.0015~\text{GeV}^2$~\cite{COMPASS:2014eqi}.
The resulting differential cross section is:
\begin{align}
    \frac{{\rm d}\sigma_\pi^\text{Born+Pol}}{{\rm d}E_\gamma}&=\frac{4 Z^2 \alpha^3}{m_\pi^2}\frac{E_\pi'}{E_\pi E_\gamma}\left[\frac{2}{3}\ln\frac{|\vec{q}_\text{max}|^2}{|\vec{q}_\text{min}|^2}-\frac{19}{9}+4\frac{|\vec{q}_\text{min}|}{|\vec{q}_\text{max}|}\right]\notag\\
    &+\frac{4 Z^2 \alpha^3}{m_\pi^2}\frac{E_\gamma}{E_\pi^2}\frac{\beta_\pi m_\pi^3}{\alpha}\left[\ln\frac{|\vec{q}_\text{max}|^2}{|\vec{q}_\text{min}|^2}-3+4\frac{|\vec{q}_\text{min}|}{|\vec{q}_\text{max}|}\right],
\end{align}
where the first term corresponds to the scattering of a point-like charged pion (the Born term), while the second term accounts for the pion structure via its electric ($\alpha_\pi$) and magnetic ($\beta_\pi$) polarizabilities.
Within Chiral Perturbation Theory (ChPT), these polarizability effects arise primarily from chiral loop corrections~\cite{Gasser:2006qa}.
Note that this expression incorporates the chiral constraint $\alpha_\pi+\beta_\pi \approx 0$.
By imposing this fundamental symmetry constraint, the structural dependence of the bremsstrahlung cross section is elegantly reduced to a single parameter governed by low-energy QCD dynamics.
While the measurement of the charged pion polarizability was the primary goal of the original COMPASS analysis~\cite{COMPASS:2014eqi}, here we adopt the theoretical prediction from ChPT as a fixed input.
The ChPT prediction, which is in excellent agreement with the value extracted from the measured COMPASS data under the SM-only hypothesis~\cite{COMPASS:2014eqi}, is given by~\cite{Gasser:2006qa}:
\begin{align}
    \beta_\pi=-(2.9\pm0.5)\times10^{-4}~\text{fm}^3.
\label{eq:betavalue}
\end{align}

\section{Primakoff ALP production}
\label{sect:alp}
The effective Lagrangian describing the ALP and its coupling to photons is given by~\cite{Georgi:1986df}:
\begin{align}
    \mathcal{L} \supset \frac{1}{2}\partial_\mu a \partial^\mu a - \frac{1}{2} m_a^2 a^2 - \frac{1}{4}g_{a\gamma\gamma}a F^{\mu \nu} \tilde{F}_{\mu\nu},
\end{align}
where $F_{\mu\nu}$ is the electromagnetic field strength tensor, $\tilde{F}_{\mu\nu}$ is its dual, $m_a$ is the ALP mass, and $g_{a\gamma\gamma}$ is the effective coupling constant.
The corresponding decay width into two photons is:
\begin{align}
    \Gamma_{a\to\gamma\gamma}=\frac{g_{a\gamma\gamma}^2 m_a^3}{64\pi},
\end{align}
which leads to a decay length in the laboratory frame of:
\begin{align}
    l_{a\to\gamma\gamma}=\frac{64\pi E_a}{g_{a\gamma\gamma}^2 m_a^4},
    \label{eq:decaylength}
\end{align}
where $E_a$ is the ALP energy.

\begin{figure}[h]
    \centering
    \includegraphics{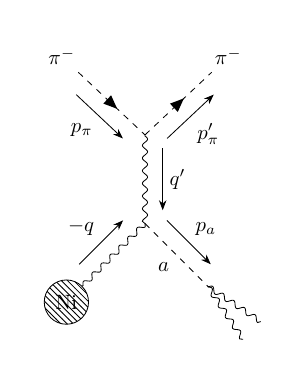} 
    \caption{Primakoff ALP production in $\pi^- \text{Ni}$ scattering, with same convention as in Fig.~\ref{fig:compton}}
    \label{fig:alp}
\end{figure}

The Primakoff production of ALPs in the scattering of high energy charged particles off a nucleus is illustrated in Fig.~\ref{fig:alp} for the pion case.
We calculate the production cross section for a given beam type $i$ using EPA.
The differential cross section with respect to the ALP energy $E_a$ is expressed as~\cite{Oliveira:2021qtq}:
\begin{align}
    \frac{{\rm d}\sigma_i^\text{ALP}}{{\rm d}E_a} = \int_{\omega'_\text{min}}^{\omega'_\text{max}} {\rm d}\omega' \frac{{\rm d}N_i}{{\rm d}\omega'} \frac{{\rm d}\sigma_{\gamma A \to a A}}{{\rm d}E_a},
    \label{eq:alp}
\end{align}
where $\omega'$ is the energy of the emitted virtual photon (with momentum $q'$) from the beam particle.
The lower integration limit, $\omega'_\text{min}$, is determined by the kinematic threshold required to produce a massive ALP.
While neglecting nuclear recoil implies $\omega'_\text{min} \approx E_a$, the exact momentum transfer constraint imposes a strict lower bound on the virtual photon energy:
\begin{align}
    \omega' - E_a \ge \frac{m_a^4}{8 m_A \omega'^2},
    \label{eq:kinematics}
\end{align}
where $m_A$ is the nucleus mass.
In addition, the maximum energy emitted from the beam particle is given by $\omega'_\text{max}=|\vec{q}_\text{max}|^2/(2m_A)+E_a$, where $|\vec{q}_\text{max}|^2$ is the maximum momentum transfer squared to the nucleus (associated with $q$), fixed to $0.0015~\text{GeV}^2$ by the experimental selection as discussed in Sec.~\ref{sect:compton}.

The virtual photon flux $dN_i/d\omega'$ emitted by the beam particle is given by~\cite{Budnev:1975poe}:
\begin{align}
    \frac{{\rm d}N_\mu}{{\rm d}\omega'} &= \frac{\alpha}{\pi \omega'} \left[ \left(1-\frac{\omega'}{E_\mu}+\frac{\omega'^2}{2E_\mu^2}\right) \ln\frac{|\vec{q}\,'_\text{max}|^2}{|\vec{q}\,'_\text{min}|^2} - \left(1-\frac{\omega'}{E_\mu}\right) \left(1-\frac{|\vec{q}\,'_\text{min}|^2}{|\vec{q}\,'_\text{max}|^2}\right) \right], \\
    \frac{{\rm d}N_\pi}{{\rm d}\omega'} &= \frac{\alpha}{\pi \omega'} \left(1-\frac{\omega'}{E_\pi}\right) \left[ \ln\frac{|\vec{q}\,'_\text{max}|^2}{|\vec{q}\,'_\text{min}|^2} - 1 + \frac{|\vec{q}\,'_\text{min}|^2}{|\vec{q}\,'_\text{max}|^2} \right].
\end{align}
These expressions are derived under the small-scattering-angle approximation. This condition is satisfied in the present analysis, as the experimental data restrict the transverse momentum of the scattered beam particles to $p_T \lesssim 0.25~\text{GeV}$~\cite{COMPASS:2014eqi}. For an incident beam energy of $190~\text{GeV}$, this corresponds to a maximum scattering angle of $\theta_\text{max} \approx 1.3~\text{mrad}$, which is within the validity regime of the approximation.

In the relevant kinematic regime, where $E_i$ denotes the incident beam energy, the minimum and maximum three-momentum transfer squared from the beam particle are given by:
\begin{align}
    |\vec{q}\,'_\text{min}|^2 = \frac{\omega'^2 m_i^2}{E_i(E_i-\omega')},\quad |\vec{q}\,'_\text{max}|^2 =4E_i(E_i-\omega').
\end{align}

For the pion flux, we adopt the point-like particle approximation within the EPA framework.
This approach ensures consistency with the Born term of the SM calculation presented in Sec.~\ref{sect:pion}, allowing systematic uncertainties associated with the equivalent photon flux modeling to largely cancel in the cross section ratio introduced in Sec.~\ref{sect:analysis}.
Furthermore, while pion polarizability is a crucial correction for the SM Compton scattering background, analogous loop-induced structure effects in the ALP production amplitude are negligible.
These structural corrections arise only at higher loop orders and are inherently heavily suppressed, rendering their contribution entirely negligible relative to the leading-order Primakoff amplitude within the currently accessible experimental sensitivity.
Following Ref.~\cite{Dusaev:2020gxi}, the Primakoff cross section for the subprocess $\gamma \text{Ni} \to a \text{Ni}$ on a point-like nucleus is:
\begin{align}
    \frac{{\rm d}\sigma_{\gamma \text{Ni} \to a \text{Ni}}}{{\rm d}E_a}
    &=
    \frac{\alpha g_{a\gamma\gamma}^2 Z^2}{8}
    \Biggl[
    \frac{E_a^2}{\omega'^2 (\omega'-E_a)}
    -\frac{m_a^4}{8\omega'^2 m_A(\omega'-E_a)^2}\notag\\
    &-\frac{m_a^2}{2\omega'^2(\omega'-E_a)}
    -\frac{m_A}{2\omega'^2}
    -\frac{E_a m_a^2}{2\omega'^2 m_A(\omega'-E_a)}
    +\frac{E_a}{\omega'^2}
    -\frac{m_a^2}{2\omega'^2 m_A}
    \Biggr].
\end{align}

\section{ALP contamination in Compton Scattering}
\label{sect:analysis}
The original COMPASS analysis reported its findings as the ratio of the observed event yield to the Monte Carlo simulation for a point-like particle, referred to as the Born term.
This ratio is defined for a beam of type $i$ as a function of the energy fraction $x = E_\text{cluster}/E_i$:
\begin{align}
    R_i^\text{obs}(x) = \frac{N_i^\text{obs}(x)}{N_i^{\text{Born}}(x)},
\end{align}
where $E_\text{cluster}$ corresponds to the energy deposited in the electromagnetic calorimeter (ECAL2) by the reconstructed photon candidate.
To reinterpret these results, we model the theoretical ratio by including the expected ALP contribution.
An ALP produced in the target travels a distance $l$ before decaying into two photons.
The distribution of path lengths is governed by the mean decay length $l_{a\to\gamma\gamma}$, defined in Eq.~(\ref{eq:decaylength}).
Since the original COMPASS analysis selected events with exactly one reconstructed cluster, an ALP signal contributes only if the two decay photons are reconstructed as a single merged cluster.
The theoretical ratio for beam type $i$, including the ALP contribution and (for the pion case) the polarizability effect, is expressed as:
\begin{align}
    R^\text{theory}_i(x) = 1 + \frac{{\rm d}\sigma_i^\text{Pol}(x)}{{\rm d}\sigma_i^\text{Born}(x)} + \frac{{\rm d}\sigma_i^\text{ALP}(x)}{{\rm d}\sigma_i^\text{Born}(x)} \times \mathcal{P}_\text{acc},
    \label{eq:R_theory}
\end{align}
where $\mathcal{P}_\text{acc}$ represents the total acceptance probability for an ALP to pass the single-photon selection.
Note that the SM background and the ALP signal are added incoherently in Eq.~(\ref{eq:R_theory}).
Quantum interference between the two processes strictly vanishes because their final states are fundamentally distinguishable in the detector framework: SM Compton scattering yields a single photon, whereas the ALP decay produces a two-photon state, regardless of the detector's finite resolving power.
The total acceptance probability, $\mathcal{P}_\text{acc}$, can be factorized as:
\begin{align}
    \mathcal{P}_\text{acc} = P_\text{decay} \times P_\text{merge}\times \frac{P_\text{survive}^\text{ALP}}{P_\text{survive}^\text{SM}}.
    \label{eq:acc}
\end{align}
The first term, $P_\text{decay}$, is the probability that the ALP decays upstream of ECAL2, located at $l_\text{ECAL2} = 33.98~\text{m}$ from the target~\cite{COMPASS:2014eqi}, and is given by:
\begin{align}
    P_\text{decay}(l_\text{ECAL2}) = 1 - \exp\left(- l_\text{ECAL2} /l_{a\to\gamma\gamma}\right).
\end{align}

The second term, $P_\text{merge}$, represents the effective probability that the two decay photons are reconstructed as a single cluster.
This signature can arise from two kinematically exclusive scenarios: highly asymmetric energy sharing and geometrical merging.

First, an event effectively mimics a single-photon topology if the ALP decay is highly asymmetric in the laboratory frame.
Due to the large Lorentz boost, a photon emitted backwards (i.e., with a decay angle $\cos\theta^* < 0$ in the ALP rest frame) is shifted to a significantly lower energy in the laboratory frame.
If this softer photon's energy falls below the ECAL2 reconstruction threshold of $E_\text{th} = 2~\text{GeV}$, it remains undetected, leaving only the forward, highly energetic photon to be reconstructed.
Applying a Lorentz boost to the laboratory frame, the energy of the softer photon is $E_{\min} = \frac{E_a}{2}(1 - \beta_a |\cos\theta^*|)$, where $\beta_a = \sqrt{1 - m_a^2/E_a^2}$.
The condition $E_{\min} < E_\text{th}$ defines a critical angular threshold, $\cos\theta^*_c$:
\begin{align}
    |\cos\theta^*| > \cos\theta^*_c \equiv \frac{1}{\beta_a}\left(1 - \frac{2 E_\text{th}}{E_a}\right).
\end{align}
Since the scalar ALP decays isotropically, $\cos\theta^*$ is uniformly distributed.
Thus, the probability of an asymmetric, single-photon topology is $P_\text{asym} = 1 - \cos\theta^*_c$.
In such asymmetric scenarios, the reconstructed cluster energy underestimates the true ALP energy by at most $2~\text{GeV}$.
However, since the experimental selection requires $E_\text{cluster} \ge 76~\text{GeV}$ ($x > 0.4$), this fractional energy shift is $\lesssim 2.6\%$, which is negligible compared to the calorimeter energy resolution and the experimental bin widths.

Second, for the remaining phase space region, occurring with a probability of $(1 - P_\text{asym})$, both photons are energetic enough to be detected. 
The event is accepted only if the two photons geometrically merge in the electromagnetic calorimeter. 
The laboratory opening angle $\theta$ reaches its minimum value $\theta_{\min} \approx 2m_a/E_a$ for perfectly symmetric decays ($\cos\theta^* = 0$). 
For other kinematics, the opening angle increases as $\theta \approx \theta_{\min}/\sin\theta^*$, which geometrically suppresses the merging probability. 
For an isotropic decay, the phase-space distribution of the laboratory opening angle is highly peaked at its minimum value. 
Since the merged acceptance is therefore overwhelmingly dominated by these minimum-angle configurations, we simplify the complex angular integration by adopting $\theta_{\min}$ to define an effective characteristic transverse separation distance of the photons at the calorimeter face, $d_\text{cluster} \approx \theta_{\min}  \left(l_\text{ECAL2}-l_{a\to\gamma\gamma}\right)$.

To quantify this merging probability, a dedicated Monte Carlo simulation of the ECAL2 cellular geometry was performed. 
Pairs of photons with this characteristic separation $d_\text{cluster}$ were uniformly projected onto the calorimeter grid. 
A strict geometrical merging condition was applied, assuming that two showers cannot be resolved by the clustering algorithm if their impact points fall within the same calorimeter cell of size $d_\text{cell}=3.8~\text{cm}$.  
Although the Moliere radius of the calorimeter material suggests that photon showers striking adjacent cells could also merge, adopting this strict single-cell geometric requirement effectively compensates for evaluating the separation at its minimum, providing a robust and conservative effective model for the merging probability.
This analytical approach captures the essential geometric acceptance while bypassing the full GEANT4 detector simulations and clustering algorithms for a dense scan over the continuous ALP parameter space. 
By evaluating the fraction of merged events as a function of the dimensionless separation parameter $z = d_\text{cluster}/d_\text{cell}$, we model this geometrical merging probability, $P_\text{geom}$, with a sigmoid function:
\begin{align}
    P_\text{geom}(z) = \frac{1}{2} \left[ 1 - \tanh\left( \frac{z - z_0}{\delta} \right) \right].
\end{align}
The shape parameters are mildly mass-dependent; however, they are well-characterized by representative values of $z_0 \approx 0.45$ and $\delta \approx 0.45$, which were extracted from the simulation fit to reproduce the cluster reconstruction efficiency. 
Therefore, the total effective merging acceptance is the statistical sum of the expectations from these two mutually exclusive regimes:
\begin{align}
    P_\text{merge} = P_\text{asym} + P_\text{geom} (1 - P_\text{asym}).
\label{eq:merging}
\end{align}

Finally, the third term in $\mathcal{P}_\text{acc}$ accounts for the relative probability of photon survival against $e^+e^-$ pair conversion in the experimental setup.
Since the analysis is based on the ratio to the SM Compton background, we must correct for the different material attenuation experienced by the single-photon background and the diphoton signal.
The survival probability of a high energy photon traversing a material thickness $x$ is determined by the pair production mean free path, $9/7~X_0$~\cite{ParticleDataGroup:2024cfk}, yielding a survival probability of $\exp(-7/9~t)$, where $t = x/X_0$ is the thickness in units of radiation lengths.

According to the experimental setup, the nickel target has a total thickness of $t_{\text{target}} = 0.285 X_0$~\cite{COMPASS:2014eqi}. Additionally, based on the internal material map of the spectrometer, the downstream material amounts to an effective thickness of $t_{\text{spec}} \approx 0.026 X_0$~\cite{COMPASS:2014eqi}.
The SM photon is assumed to be produced, on average, at the center of the nickel target. It therefore traverses half of the target's material plus the downstream components. Its survival probability is:
\begin{align}
    P_\text{survive}^\text{SM} = \exp\left[-\frac{7}{9} \left(\frac{t_{\text{target}}}{2} + t_{\text{spec}}\right)\right].
\end{align}

In contrast, the ALP travels a distance before decaying into two photons. To maintain analytical consistency with the geometric merging model, we evaluate this material penalty at the mean decay length, $l_{a\to\gamma\gamma}$.
Assuming the ALP is also produced at the center of the target, the two resulting photons must traverse the remaining fraction of the target, if any, plus the downstream spectrometer.
The survival probability for the diphoton state is therefore:
\begin{align}
    P_\text{survive}^\text{ALP} = \exp\left\{-\frac{7}{9} \times 2 \left[ \frac{t_{\text{target}}}{l_{\text{target}}} \left(\frac{l_{\text{target}}}{2} - l_{a\to\gamma\gamma}\right) \Theta\left(\frac{l_{\text{target}}}{2} - l_{a\to\gamma\gamma}\right) + t_{\text{spec}} \right]\right\},
\end{align}
where $\Theta(x)$ denotes the Heaviside step function, and $l_\text{target}=4.2$ mm is the thickness of the target~\cite{COMPASS:2014eqi}. In this formulation, we conservatively apply the full downstream spectrometer material penalty, $t_\text{spec}$, to the diphoton state regardless of the decay length. This effectively neglects the kinematic scenario where long-lived ALPs decay within or downstream of the spectrometer components, thereby bypassing some material attenuation. Given that $t_\text{spec}$ is small relative to the effective target thickness, this simplification yields a negligible impact on the final exclusion limits. 
This relative material correction, $P_\text{survive}^\text{ALP}/P_\text{survive}^\text{SM}$, introduces a non-trivial kinematic dependence: long-lived ALPs decaying outside the target experience less material suppression than the SM background, yielding a ratio greater than unity, whereas short-lived ALPs decaying promptly inside the target suffer a severe diphoton penalty, reducing the ratio below one.

To illustrate the relative importance of the factors defining the total acceptance probability in Eq.~\ref{eq:acc}, we can identify distinct kinematic regimes where different terms dominate. In the most sensitive region of the parameter space, $\mathcal{P}_\text{acc}$ reaches $\mathcal{O}(1)$. However, at low masses and weak couplings, the ALPs become long-lived, and the decay probability $P_\text{decay}$ becomes the dominant source of signal suppression. In the parameter space, the boundary where $P_\text{decay}$ drops below $10^{-2}$ is strictly dictated by the decay length kinematics and follows the contour $g_{a\gamma\gamma} m_a^2 \approx 10^{-7}~\text{GeV}$; below this threshold, the acceptance vanishes rapidly. Conversely, at higher masses, i.e., $m_a \gtrsim 30$ MeV, the larger opening angle between the decay photons causes the geometrical merging probability to vanish. Here, $P_\text{merge}$ plummets below $10^{-2}$ and dominates the acceptance loss, leaving the highly asymmetric decays, $P_\text{asym}$, as the sole source of residual acceptance. Finally, the relative material correction factor, $P_\text{survive}^\text{ALP}/P_\text{survive}^\text{SM}$, generally acts as an $\mathcal{O}(1)$ modulation across the entire parameter space; it introduces a mild penalty for promptly decaying ALPs and a slight enhancement for long-lived ones, but it does not drive the overall exclusion boundaries.

\begin{figure}[H]
    \centering
    \includegraphics[width=0.49\linewidth]{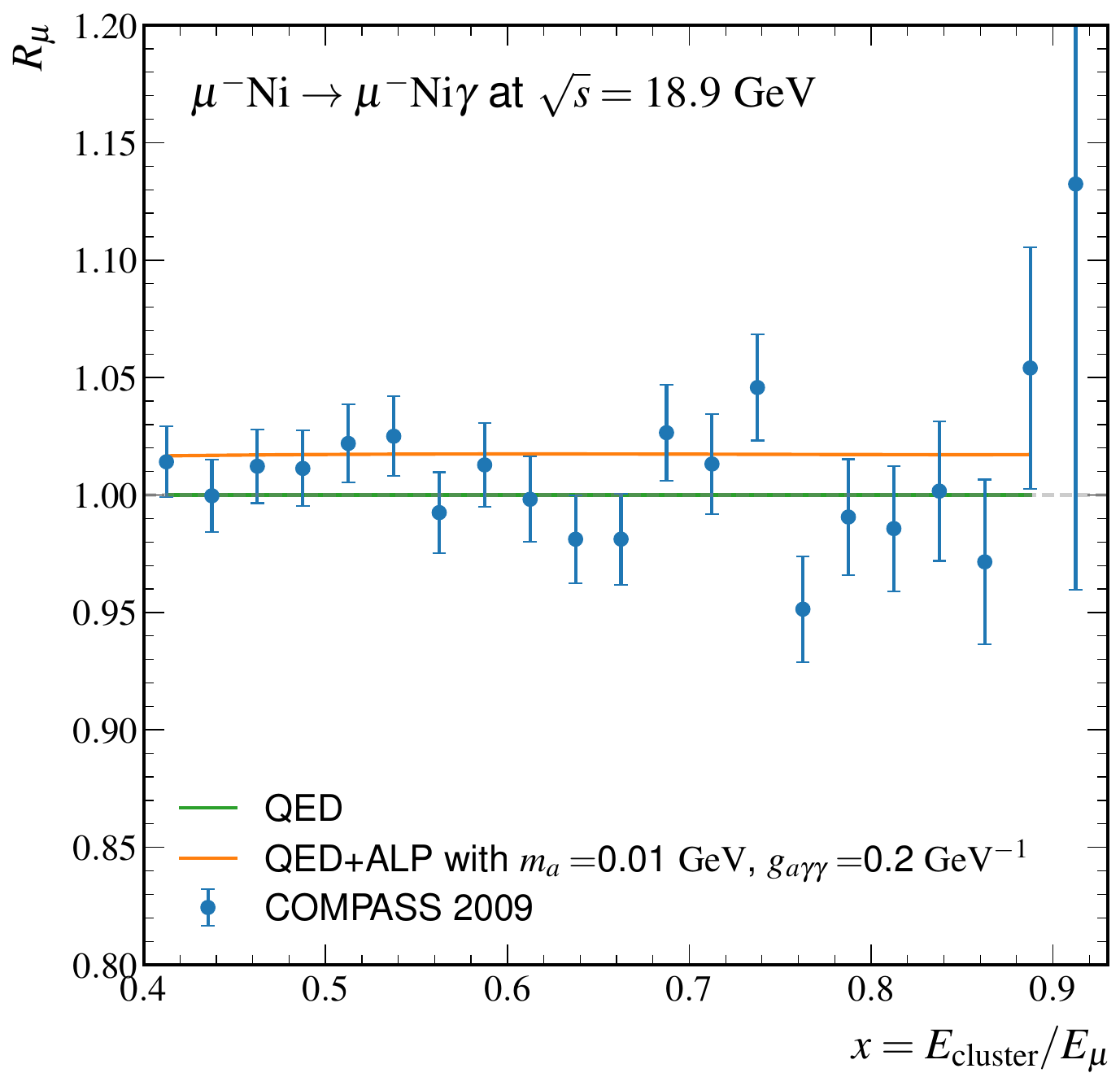}
    \includegraphics[width=0.49\linewidth]{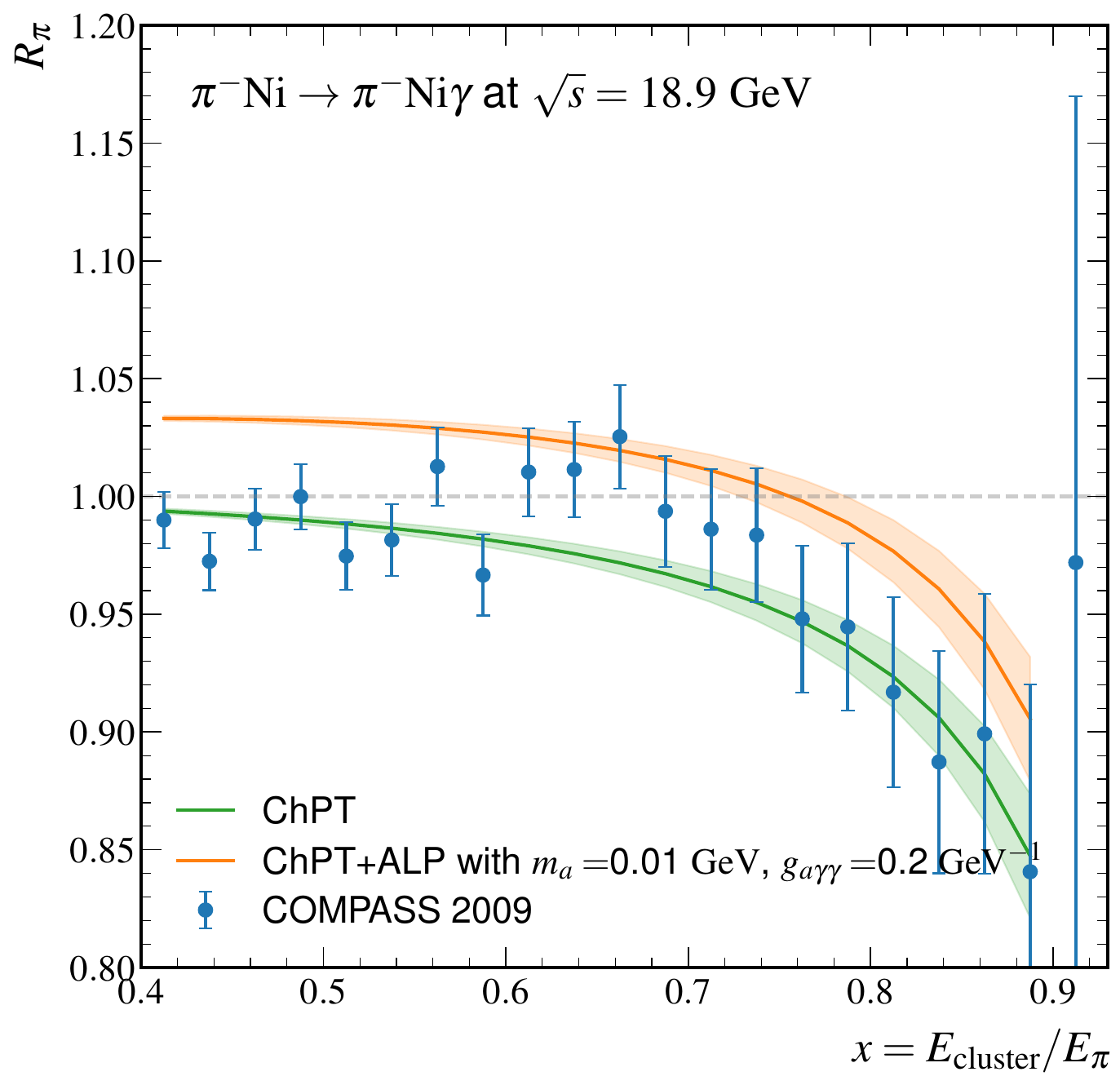}
    \caption{Comparison of the measured ratio $R_i$ with the SM expectation.
    The orange curves illustrate the expected deviation for benchmark ALP signals with representative parameters.
    For the pion beam case, the theoretical prediction incorporates the ChPT result for charged pion polarizability. Error bars represent the total experimental uncertainty, calculated as the quadratic sum of the statistical and systematic uncertainties reported by the Ref.~\cite{COMPASS:2014eqi}}
    \label{fig:ratio}
\end{figure}

The experimental ratios $R_i^\text{obs}$ and their associated uncertainties were extracted from Ref.~\cite{COMPASS:2014eqi}.
Figure~\ref{fig:ratio} illustrates the data alongside the expected deviation for a benchmark ALP signal and the ChPT prediction for the pion beam case.

To derive quantitative limits while properly accounting for the bin-to-bin correlations of the shape-dependent systematic uncertainties, we employ a profile likelihood approach using a nuisance parameter. The $\chi^2$ function for each beam type is defined as:
\begin{align}
    \chi_i^2(\alpha_i) = \sum_k \frac{\left( R_{i,k}^\text{obs} - \left( R_{i,k}^\text{theory} + \alpha_i \sigma_{i,k}^\text{syst} \right) \right)^2}{(\sigma_{i,k}^\text{stat})^2 + (\sigma_{i,k}^\text{theory})^2} + \alpha_i^2,
    \label{eq:chi2}
\end{align}
where $k$ denotes the kinematic bins in the range $0.4 < x < 0.9$, and $\alpha_i$ is a continuous parameter that shifts the theoretical prediction within the fully correlated systematic uncertainties. 
Here, $\sigma_{i,k}^\text{stat}$ and $\sigma_{i,k}^\text{syst}$ denote the statistical and systematic uncertainties for the $k$-th bin, respectively, as reported separately in Ref.~\cite{COMPASS:2014eqi}. For each point in the parameter space, the function is analytically minimized with respect to $\alpha_i$ to obtain the final profiled $\chi_i^2$. 
For the pion case, $\sigma_{i,k}^\text{theory}$ accounts for the uncorrelated uncertainty in the ChPT prediction of the polarizability according to Eq.~(\ref{eq:betavalue}), while for the muon case, theoretical uncertainties are negligible.

\section{Results and conclusion}
\label{sect:results}
\begin{figure}[H]
    \centering
    \includegraphics[width=0.98\linewidth]{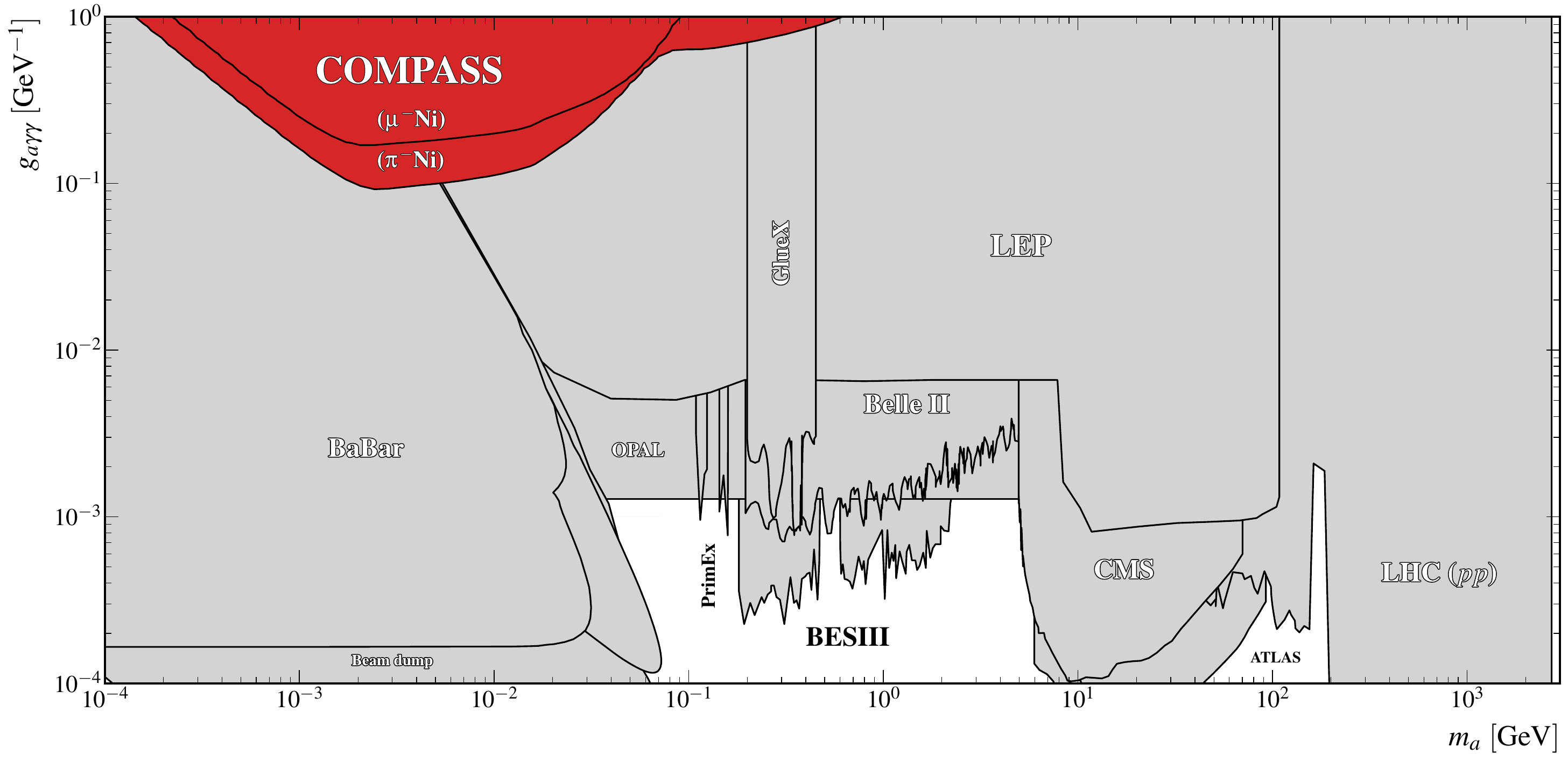}
    \caption{The 95\% C.L. exclusion limits on the ALP-photon coupling as a function of ALP mass, derived from the COMPASS 2009 data, compared with existing constraints~\mbox{\cite{CHARM:1985anb, Riordan:1987aw, Blumlein:1990ay, PrimEx:2010fvg, Jaeckel:2015jla, Knapen:2016moh, Dolan:2017osp, CMS:2018erd, Aloni:2019ruo, NA64:2020qwq, ATLAS:2020hii, Belle-II:2020jti, BESIII:2022rzz, Pybus:2023yex, BESIII:2024hdv, FASER:2024bbl}} compiled in Ref.}~\cite{AxionLimits}.
    \label{fig:exclusion}
\end{figure}

For both the muon and pion beam datasets, we determine the exclusion limits at $95\%$~C.L. using a grid scan approach. 
For each fixed ALP mass $m_a$, we vary the coupling $g_{a\gamma\gamma}$ and calculate the difference $\Delta \chi_i^2 = \chi_i^2 - \chi_i^2(g_{a\gamma\gamma}=0)$, where $\chi_i^2(g_{a\gamma\gamma}=0)$ corresponds to the SM-only hypothesis.
The exclusion boundary is set at the coupling value where $\Delta \chi_i^2 \geq 3.84$, corresponding to a $\chi^2$ distribution with one degree of freedom.
The resulting exclusion limits are presented in Fig.~\ref{fig:exclusion}, alongside existing constraints from other experiments~\mbox{\cite{CHARM:1985anb, Riordan:1987aw, Blumlein:1990ay, PrimEx:2010fvg, Jaeckel:2015jla, Knapen:2016moh, Dolan:2017osp, CMS:2018erd, Aloni:2019ruo, NA64:2020qwq, ATLAS:2020hii, Belle-II:2020jti, BESIII:2022rzz, Pybus:2023yex, BESIII:2024hdv, FASER:2024bbl}}, as compiled in Ref.~\cite{AxionLimits}.

It is worth noting that a statistically combined global fit of both datasets yields no discernible improvement over the pion-only case, as the $\pi^-{\rm Ni}$ data overwhelmingly dominates the sensitivity across the entire probed mass range. Therefore, the exclusion limits for the two beams are presented independently.

Several distinct features of these limits warrant discussion.
First, the constraints derived from the pion beam are significantly stronger than those from the muon beam.
Remarkably, this enhancement is observed despite the comparable integrated luminosities and statistics of the final event samples~\cite{COMPASS:2014eqi}.
This difference is primarily kinematically driven: the SM bremsstrahlung cross section scales as $1/m_i^2$.
Consequently, the heavier pion mass naturally suppresses the SM Compton background relative to the muon case, leading to a much larger expected deviation from the SM hypothesis.

The shape of the excluded region is strictly determined by the interplay between the ALP decay kinematics and its lifetime.
In the low-mass regime, the sensitivity is limited by the long decay length $l_{a\to\gamma\gamma}$.
Light ALPs become long-lived and typically traverse the detector volume without decaying, exponentially suppressing the probability of observing a signal inside the fiducial volume.
Conversely, at higher masses, the transverse spatial separation of the photons generally exceeds the calorimeter cell size, leading to a vanishing geometrical merging probability, $P_\text{geom}$.
However, the asymmetric decay contribution, $P_\text{asym}$, remains finite, sustaining a small but non-zero effective acceptance.

This asymmetric effect creates the distinct high-mass tail observed in the pion exclusion limits, extending the sensitivity up to $\sim 600~\text{MeV}$, albeit at stronger couplings due to the reduced acceptance and the diminished Primakoff cross section.
While this mechanism applies equally to both beam types, the corresponding tail is absent in the muon limit within the probed coupling range.
This absence is a direct consequence of the significantly larger SM muon bremsstrahlung background;
in this reduced-acceptance regime, the heavily suppressed signal yield is entirely overwhelmed by the background, pushing the required coupling for exclusion beyond the plotted range.

As illustrated in Fig.~\ref{fig:exclusion}, the limits derived in this work effectively span the intermediate region between low-mass beam dump experiments and high energy collider searches.
It is worth noting that at the time of the original COMPASS data publication, this parameter space was largely unexplored.
While subsequent reinterpretations of LEP data~\cite{Jaeckel:2015jla,Knapen:2016moh} have since constrained this region, the COMPASS results provide an independent and complementary exclusion leveraging a completely different production mechanism.

Looking forward, the reinterpretation framework established in this work holds significant potential for broader applications. Specifically, our single-cluster analysis can be directly applied to the larger dataset collected by COMPASS in 2012 with an identical setup~\cite{Friedrich:2016gqb}. Furthermore, as the COMPASS facility has been upgraded to the AMBER experiment~\cite{Adams:2018pwt}, this methodology remains highly relevant for its upcoming physics program. AMBER’s proposed measurement of the kaon polarizability via the analogous Primakoff reaction, $K^- \text{Ni} \to K^- \text{Ni}\gamma$, will utilize a $100~\text{GeV}$ kaon beam and a remarkably similar spectrometer configuration. The heavier initial-state kaon will modify the Standard Model bremsstrahlung background, providing a distinct and compelling avenue to further explore the ALP parameter space using the exact unresolved diphoton framework developed here.

Beyond this specific regime, there are also vital opportunities to probe ALPs through complementary kinematics. For instance, the COMPASS collaboration is currently analyzing Primakoff $\pi^0$ production using the same 2009 dataset~\cite{Ecker:2023qae}. Because this $\pi^0$ analysis explicitly reconstructs resolved diphoton final states, the strict merging constraint required in our current work is naturally relaxed. Consequently, such an approach would complement our results by potentially extending the sensitivity to higher ALP masses, where the decay photons are well-separated in the calorimeter.

\section*{Data Availability Statement}
The datasets generated and analyzed during the current study are available from the corresponding author upon reasonable request.

\acknowledgments
The author thanks the COMPASS collaboration for clarifying the supplementary material for Ref.~\cite{COMPASS:2014eqi}.
Grateful acknowledgment is also given to Jan Matousek and Sahar Safari for useful discussions.

\bibliography{biblio}

\end{document}